# Precision and Accuracy Testing of FMCW Ladar Based Length Metrology

Ana Baselga Mateo[1], Zeb W. Barber[1*]

[1]Spectrum Lab, Montana State University,
Bozeman, MT 59717
*Corresponding author: barber@spectrum.montana.edu

The calibration and traceability of high resolution frequency modulated continuous wave (FMCW) ladar sources is a requirement for their use in length and volume metrology. We report the calibration of a FMCW ladar length measurement system by use of spectroscopy of molecular frequency references HCN (C-band) or CO (L-band) to calibrate the chirp rate of the FMCW source. Propagating the stated uncertainties from the molecular calibrations provided by NIST and measurement errors provides an estimated uncertainty of a few ppm for the FMCW system. As a test of this calibration, a displacement measurement interferometer with a laser wavelength close to that of our FMCW system was built to make comparisons of the relative precision and accuracy. The comparisons performed show ppm agreement which is within the combined estimated uncertainties of the FMCW system and interferometer.

## Introduction

Frequency modulated continuous wave (FMCW) ladar (also known as swept frequency/wavelength interferometry [1]) is a well-known time-of-flight distance measurement technique, which as a consequence of the stretched processing technique has several advantages for ladar based metrology applications. Enabled by the coherent heterodyne detection, stretched processing accesses the fine temporal resolution provided by the full bandwidth of the chirp (as large as several THz) with only low bandwidth detectors and analog-to-digital converters (ADC's). Additionally, the coherent detection technique provides high sensitivity to low power returning signals and large dynamic range [2]. This relieves the need for cooperative targets such as mirrors or retro-reflectors. Moreover, spreading the pulse in time also lowers the peak optical powers, making it more compatible with fiber optic amplification and delivery. These benefits have made FMCW ladar a promising solution for many ranging systems in addition to non-contact and stand-off metrology [3,4].

FMCW systems have been proposed in the past to sense both absolute and relative lengths on the micro to nanometer range [1,5,6]. Precise linearization of ultrabroadband (several THz) optical frequency chirps in a FMCW ladar configuration has led to results with resolution of 31 µm [7,8]. This ability to resolve extremely small ranges makes FMCW ladar an attractive system for high accuracy applications. As the detected beat frequency is proportional to the desired range measurement through the speed of light and the chirp rate, the latter is the main parameter that must be calibrated in an accurate FMCW ladar system. Although characterization against an optical frequency comb showed that the sweep can be linearized and calibrated to less than 15 ppb [9,10,3], ~ppm accuracy would be sufficient for most applications and a less expensive, traceable calibration method is highly desired. Therefore, we focus in this paper on calibration by use of molecular frequency references HCN (C-band) or CO (L-band). The absorption frequencies of these molecular gases have been calibrated and certified by NIST [11,12] and can act as a "frequency ruler" to calibrate the chirp rate of a FMCW ladar system. Furthermore, in order to demonstrate the chirp rate calibration accuracy for the FMCW ladar source translates into the accuracy of the length measurements, a head-to-head comparison against a homemade continuous wave (CW) interferometer displacement measurement system was performed.

The FMCW ladar experimental setup is shown in Fig.1. For the experimental work described in this paper, linear frequency chirps were used, and the chirp laser sources provide active laser stabilization during the frequency sweep, which dramatically reduces sweep nonlinearities and substantially increases the swept source coherence length. The time delay between optical paths introduces a frequency difference between Tx/Rx and LO paths. Coherently mixing the two optical signals with balanced coherent detection provides coherent signal amplification for shot-noise limited detection and simultaneously de-chirps the received signal. After detection and digitization by use of a NI-5122 ADC computer card, a Fast Fourier Transform (FFT) calculates the frequency profile which is scales to the range through the chirp rate, $\kappa$, and the speed of light as $R = f\kappa/(2c)$. A thick (1/2") uncoated window was placed in the free space section of the FMCW ladar path to serve as a range reference to take out range drifts due to the fiber portions of the path. The close range and collimated geometry allowed the range peak signals for the target and the reference to have carrier-to-noise ratio (CNR) exceeding 60 dB even if the reflection from the target was diffuse.

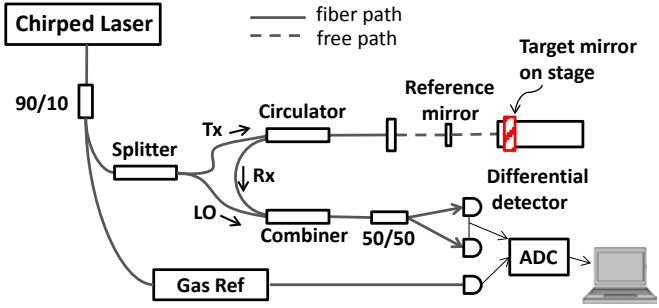

Fig. 1. FMCW ladar set up. A motorized stage was used to move the target in linear steps in both directions. The target mirror was also the target for the CW interferometer setup detailed in the following section. This allowed direct measurement comparison of the two methods to determine the FMCW ladar precision and accuracy.

For the results presented in this paper two different chirped lasers were used as the chirped source. Laser A [13] was operated in C-band (1535-1565 nm) with a chirp rate of ~227 MHz/µs and a repetition rate of about 30 Hz. Laser B [7,9] was operated in L-band (1555-1595 nm) with a chirp rate of ~5 MHz/µs and a repetition rate of about 0.5 Hz. The two lasers represent a later (A) and earlier stage (B) development of the FMCW ladar chirped sources.

### Chirp Rate Calibration

The chirp rate calibration process consists of using the relatively sharp spectral features and broad frequency coverage of the optical absorption in a molecular gas absorption cell as a traceable ruler of the optical frequency. The two gas cells used for this work were a 10 Torr $H^{13}C^{14}N$ cell for the C-band and a 100 Torr $^{12}C^{16}O$ cell for the L-band both from Wavelength References Inc. The transmission of chirped laser light through the gas cell was detected using an homemade auto-balanced detector and recorded using the second channel of the NI-5122 ADC. After recording, the absorption peaks were fit with a Voigt profile to find the center (in time) of each strong absorption feature. The measured centers were then plotted against the calibrated frequencies from the NIST certified SRM 2519a or SRM 2514 values [14,15], which were adjusted for the difference in pressure of the cell. Each absorption center frequency was assigned an uncertainty equal to the one sigma uncertainty stated in the certification (the uncertainties based on the residuals of the Voigt fits were generally negligible). Applying a weighted polynomial fit [16] to the centers returns the chirp center frequency, the chirp rate (slope $\kappa$), chirp rate curvature (if desired), and the standard uncertainties of these values based on the residuals and weights. Assuming a constant reflection with frequency, the average frequency slope (chirp rate) determines the proper scaling parameter from frequency to range. This means that care must be taken to compensate the origin dependent coupling of the linear coefficient with higher order coefficients if 2nd order or higher polynomials are used for the fit. As an additional optimization step, the cell pressure was used as a free parameter to minimize the residuals, although this optimized pressure deviated from that specified by the manufacturer by approximately 5%, was consistent from calibration to calibration, and had negligible effect on the final calibration.

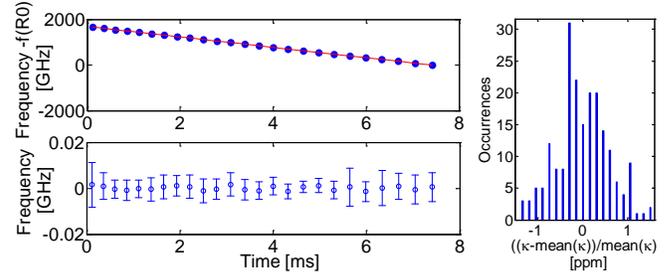

Fig. 2. (Upper Left) A fit to the measured line center times vs. certified absorption frequencies of HCN for Laser A (1535-1565 nm). (Lower Left) Residuals of fit with error bars representing certified uncertainties. (Right) A histogram of the chirp rate results for 200 calibrations with Laser A.

Fig. 2 and Fig. 3 illustrate the results of the calibration of Laser A vs. HCN and Laser B vs. CO respectively. In both cases, the high linearity of the chirp results in a very precise correspondence between the measured line center times and certified frequency values resulting in residuals from the fit, $\sigma_{HCN} = 0.93$ MHz, that are significantly smaller than the line-by-line stated uncertainties from NIST (up to several MHz). However, to ensure traceability in the chirp rate we use the latter uncertainties as weights in the fit producing a result of $\kappa$=227.2944(17) MHz/µs for the HCN calibration, which corresponds to a 7.5 ppm systematic uncertainty. This choice leads to $\chi_\nu^2$ test values of much less than one. This is also supported by the histograms in Fig. 2 and Fig. 3, which give an estimate of the statistical uncertainties in the chirp rate parameter much less than the systematic uncertainties. All this suggests that the final calibration uncertainty in the chirp rate is dominated by the certified uncertainties in the absorption line frequencies. The results from the CO calibrations using Laser B are similar to the Laser A vs. HCN calibrations, except Laser B suffered from infrequent "glitches" in the chirp stabilization lock leading to a significant tail (see histogram in Fig. 3) that caused increased uncertainties. In addition, the uncertainties for CO in SRM 2514 are large than that of SRM 2519a leading to larger systematic uncertainty (20 ppm vs. 7.5 ppm). The final calibration uncertainties for both lasers are summarized in Table 1.

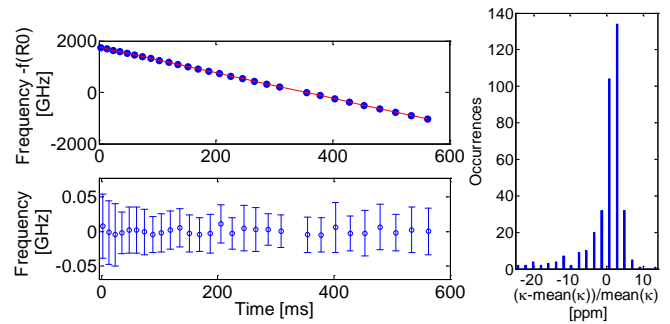

Fig. 3. (Upper Left) A fit to the measured line center times vs. certified absorption frequencies of CO for Laser B (1555-1595 nm). The resulting chirp rate is $\kappa$=4.9788(1) MHz/µs for an systematic uncertainty of 20 ppm. (Lower Left) Residuals of fit with error bars representing certified uncertainties. (Right) A histogram of the chirp rate results for 375 calibrations with Laser B.

Table 1. Chirp rate calibration results.

|  | Laser A | Laser B |
|---|---|---|
| Chirp Region [nm] | 1535-1665 | 1555-1595 |
| Molecular gas | HCN | CO |
| Systematic uncertainty | 7.5 ppm | 20 ppm |
| Statistical uncertainty | 0.6 ppm | 6 ppm |
| Combined uncertainty | 7.5 ppm | 21 ppm |

## CW Interferometer

To test that these calibrations translate into actual length measurements a displacement measuring interferometer was constructed to provide head-to-head measurements of the displacement of a mirror on a linear stage. The CW interferometer system and signal processing chain is outlined in Fig. 4.

The laser source for the interferometer was a low phase noise and narrow linewidth laser centered at 1536.475 nm as measured by a Burleigh WA-1500 wavemeter, which has a specified accuracy of 0.2 ppm. However, the current accuracy of the wavemeter was not been verified through other means. The choice of laser and wavelength was based on availability from prior unrelated work and that it is relatively close to the wavelength of the FMCW chirp lasers (as compared to a 633 nm HeNe laser). This reduced the range shift due to air dispersion in the ranging path to less than 0.1 ppm and allowed the two systems to use the same optics including the target mirror.

The interferometer system was based on a phase modulation technique designed to provide for IQ quadrature detection of the interferometer phase. Details of the interferometer can be found in Ref. [17]. This setup was constructed with a movable mirror (target) in one interferometer arm and a bulk electro-optic phase modulator (EOM) in the other arm. The EOM was driven sinusoidally at 5 MHz using a resonant tank circuit to provide modulation of the signal for synchronous lock-in detection (see Figure 4). Phase sensitive detection of both the first and second harmonic of the modulation frequency provides cosine and sine (IQ) outputs whose ratio (i.e. the tangent of the measured phase angle) is independent of laser power and fringe contrast. These outputs were sampled with an analog-to-digital-converter, and also thresholded at their zero-crossing using a comparator and fed into a digital quadrature counter (HCTL2022). The quadrature detection design allows both the sign of the displacement to be tracked and results in a discrete resolution of $\lambda/8$ (or ~192 nm) per count. For increased resolution, the intra-fringe phase of the interferometer can be calculated by using the analog outputs.

## FMCW Ladar vs. Interferometer Comparisons

For the FMCW ladar vs. interferometer comparisons the displacement of the end mirror of the Michelson interferometer was measured simultaneously with the CW interferometer and the FMCW ladar. The two interrogation paths were multiplexed on the mirror by use of knife edge mirror to bring the FMCW ladar beam in parallel to the interferometer beam. The parallelism of the two beams was much better than 1 mm over one meter, results in an uncertainty due to parallelism of less than 1 ppm. The longest lengths of the target arm for the interferometer and the free space distance between the reference and the target in the FMCW path were both slightly less than 1 m. A box was placed over most of the propagation path of both the FMCW ladar and interferometer systems to mitigate air currents that can lead to optical path length changes. First, a long series of simultaneous measurements with the mirror held stationary was performed to estimate the range precision of the FMCW measurement system and stability of the target (see Fig. 5). These measurements show a standard deviation of 0.63 μm for Laser A and 0.14 μm for Laser B. This shows that both lasers and the test system have the potential for less than 1 ppm measurements in a single shot.

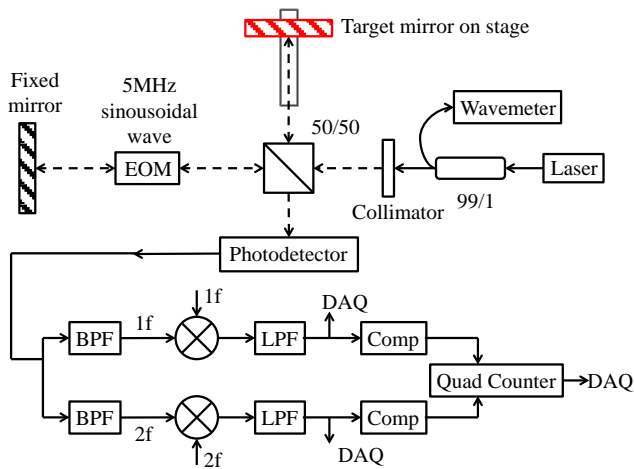

Fig. 4. Interferometer and phase sensitive detection signal processing chain. 1f = first harmonic, 2f = second harmonic, BPF = bandpass filter, LPF = lowpass filter, DAQ = Digital Acquisition Device (NI 6008), Comp = comparator, Quad = quadrature.

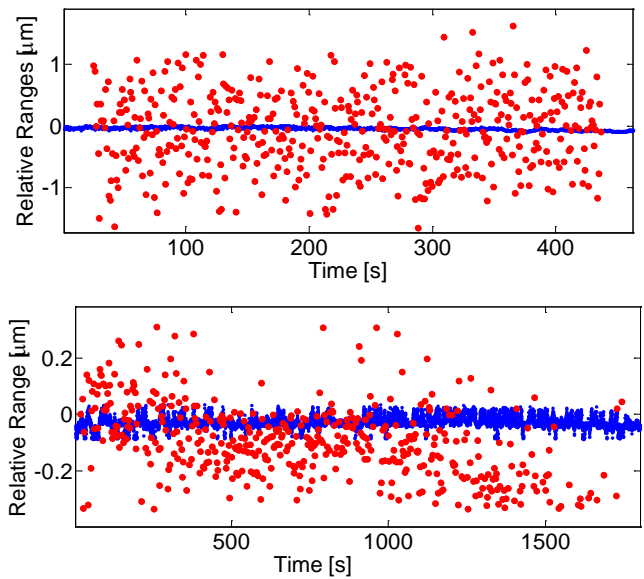

Fig. 5. Stability and precision of FMCW ladar (larger red dots) and CW interferometers measurements (small blue dots) for the mirror target for Laser A (top) and Laser B (bottom) respectively. The standard deviation of the results for interferometer is 0.02 μm, 0.63 μm for Laser A, and 0.14 μm for Laser B.

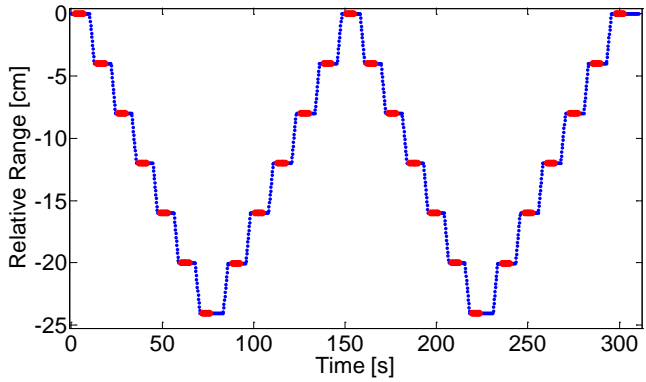

Fig. 6. Mirror target moving back and forth along 24 cm in 4 cm steps (5 ladar measurements per step). The blue points indicate the interferometer results and the red ones the ladar using Laser A.

To test the calibration accuracy the mirror target was stepped six times in increments of 4 cm for a total displacement of 24 cm, then stepped back in the same manner to the original position, then repeated (see Fig. 6). At each step position the target range was measured with the FMCW ladar system five times and the CW interferometer returned values at a 10 Hz rate. To characterize the calibration accuracy the interferometer relative range for this 24 cm target motion is plotted versus the ladar relative range and a line is fit to the points (see Fig. 7 and Fig. 8). Any deviation of the line slope from 1 indicates a systematic calibration error of the chirp laser. To find this slope, a weighted polynomial fit was used introducing the values obtained for the precision of the FMCW ladar as weights. The slope from the fit for Laser A was 0.9999925(8) and Laser B was 1.0000098(1), which represent deviations from unity of 7.5 ppm and 9.8 ppm, respectively. This compares favorably to the calibration uncertainties from the HCN and CO calibrations of 7.5 ppm and 21 ppm.

In addition to measuring mirror targets, the FMCW ladar system provides sufficient sensitivity and CNR to accurately measure distances to diffuse targets. To test this ability and the accuracy, the mirror target was translated transverse to the beams to make the FMCW ladar beam incident on the mirror mount rather than the mirror itself. Despite the mirror mount being constructed of black anodized aluminum, the diffuse reflection was sufficient. Similar measurements were repeated in this configuration and are summarized in Fig. 9 and Fig. 10. In addition to the range precision and stability for the diffuse target showing only minimal reduction (see Fig. 9) the comparison of the calibration accuracy against the CW interferometer shows no degradation (see Fig. 10). For the diffuse target configuration the stepped motion target comparison was performed for Laser A only. This resulted in an interferometer vs. ladar comparison with a slope of 0.9999932(13) or a deviation from unity of 6.8 ppm. The results of these comparisons are summarized in Table 2.

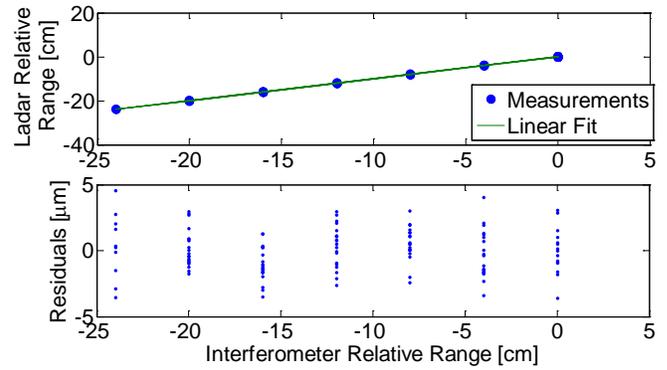

Fig. 7. Interferometer vs. Laser A relative range results for a mirror target. A linear fit of the plotted measurements gives the residuals shown in the bottom plot.

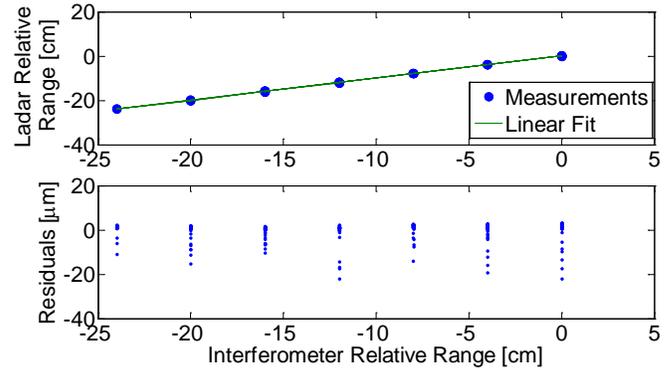

Fig. 8. Interferometer vs. Laser B relative range results for a mirror target. A linear fit of the plotted measurements gives the residuals shown in the bottom plot.

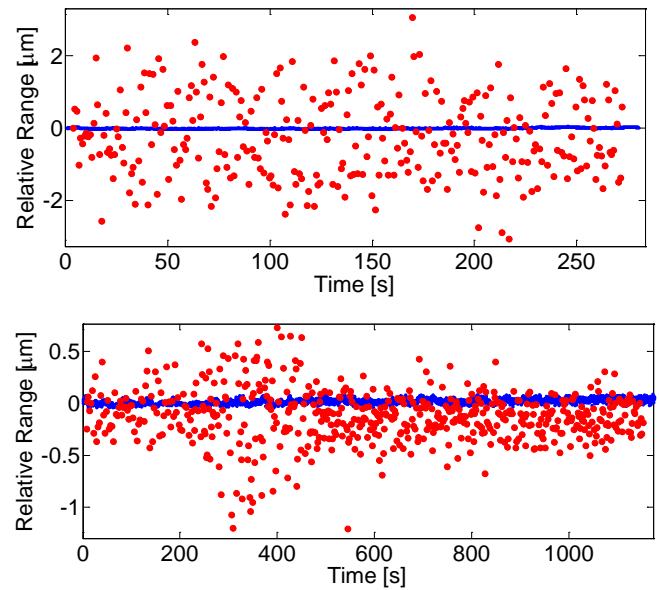

Fig. 9. Stability and precision of FMCW ladar (larger red dots) and CW interferometers measurements (small blue dots) for the diffuse target for Laser A (top) and Laser B (bottom) respectively. The standard deviation of the results was 1.12 µm for Laser A, and 0.258 µm for Laser B against the diffuse target. Laser B uses more bandwidth and has higher CNR resulting from the longer integration time resulting in more precise measurements.

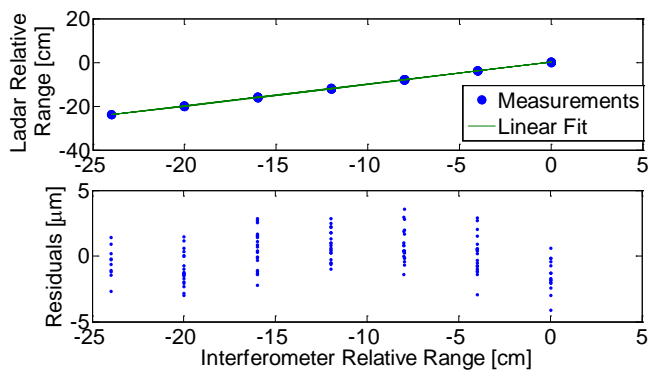

Fig. 10. Interferometer vs. Laser A relative range results for a diffuse target. A linear fit of the plotted measurements gives the residuals shown in the bottom plot.

Table 2. FMCW ladar vs. CW interferometer

|  |  | Laser A | Laser B |
|---|---|---|---|
|  | Chirp Region [nm] | 1535-1665 | 1555-1595 |
| Mirror | Measurement precision | 0.63 µm | 0.14 µm |
|  | Comparison deviation | 7.5 ppm | 9.8 ppm |
| Diffuse | Measurement precision | 1.12 µm | 0.26 µm |
|  | Comparison deviation | 6.8 ppm | - |

## Conclusion

We have demonstrated an inexpensive traceable method for calibration of FMCW ladar systems by calibrating their chirp rates against the molecular frequency references $H^{13}C^{14}N$ or $^{12}C^{16}O$. With this system we have estimated the chirp rate of two sources (Laser A and Laser B) with traceable fractional uncertainty down to a several ppm. The calibration accuracy is limited by the conservative systematic uncertainties of the absorption line centers provided by NIST. The statistical uncertainties in the calibration method can support uncertainties at the sub-ppm level. The head-to-head comparison of the FMCW ladar systems against a displacement CW interferometer show that the calibration accuracy against the molecular references translates to that of the length measurement and supports the stated calibration uncertainties. The ranging capability of the FMCW ladar system can be better than 1 ppm precision for a fixed target in air against mirror (cooperative) and diffuse (non-cooperative) targets. These precisions and accuracies are sufficient for most industrial metrology applications. Other benefits of the FMCW ladar metrology method include: removing the requirement to continuously track a target as in standard CW interferometry, the ability to range to multiple surfaces (targets) simultaneously such as multiple surfaces in transparent solids or for resolving small stepped features on surfaces, and high coherent sensitivity allowing high resolution, precision, and accuracy measurements to diffuse targets with inherent insensitivity to stray light. Finally, in a separate article we describe and demonstrate a new concept for extending the FMCW ladar system to multi-dimensional, trilateration based metrology [xx].

## Acknowledgements

The authors gratefully thank Bridger Photonics Inc. for the loan of the SLM-M laser source (i.e. Laser A). They also would like to acknowledge discussions with Randy Reibel and Mike Thorpe of Bridger Photonics that helped to resolve some earlier calibration issues. This work was supported by NSF grant MCME\GOALI #1031211.